\documentclass[aps,preprint]{revtex4}%
\usepackage{amsfonts}
\usepackage{amsmath}
\usepackage{amssymb}
\usepackage{subfigure}
\usepackage{graphicx}
\usepackage{multirow}
\usepackage{color}%
\usepackage{float}
\usepackage{cleveref}

\begin{document}
\title{Thermodynamic instabilities of Kerr-Newman Ads black holes}
\author{}
\author{Yiqian He$^{a,b}$}
\email{heyiqian@std.uestc.edu.cn}
\author{Benrong Mu$^{a,b,c}$}
\email{benrongmu@cdutcm.edu.cn}
\affiliation{$^{a}$ Center for Joint Quantum Studies, College of Medical Technology,
	Chengdu University of Traditional Chinese Medicine, Chengdu, 611137, PR China}
\affiliation{$^{b}$ School of Physics, University of Electronic Science and Technology of China, Chengdu, 611731, China}
\affiliation{$^{c}$Center for Theoretical Physics, College of Physics,
Sichuan University, Chengdu, 610064, PR China}

\begin{abstract}
In~\cite{Johnson:2019mdp}, Clifford put forward that the specific heat at constant volume $C_{V}$ is always negetive for the super-entropy black hole. Later in~\cite{Cong:2019bud}, Robert et al. found $C_{V}>0$ in certain region for the generalized exotic
BTZ black holes. Futhermore, they proposed a new conjecture that as the $C_{V}>0$, the specific heat at constant pressure $C_{P}$ always satisfies $C_{P}<0$ in super-entropy black hole. In this paper, we examine the both conjectures with regard to the thermodynamic instability of super-entropy black hole. By means of a super-entropy black hole, the Kerr-Newman-AdS black hole in a coordinate system that rotates at infinity, we detailedly analyze its character and notice there exists the region where $C_{V}>0$ and meanwhile $C_{P}>0$. Hence, we find a counterexample to the two conjectures.
\end{abstract}

\maketitle
\tableofcontents{}

\bigskip{}

\section{Introduction}
Since prominent contribution of Hawking and Page~\cite{Hawking:1982dh}, the thermodynamics of AdS space time attracts a great deal of scholars to investigate. Generally speaking, a black hole is limited by the relation $U=M$ and the first law of thermodynamics $dU=TdS$. In~\cite{Johnson:2019mdp}, the instability of a gravitational system can be expressed by the specific heat
\begin{equation}\label{eqn:Q30}
C=T\frac{\partial S}{\partial T}=-\frac{1}{8\pi T^{2}},
\end{equation}
and the negetive specific heat leads to instable system.
 In recent years, increasing studies indicate the mass of the black hole is regarded as the enthalpy instead of energy~\cite{Kastor:2009wy}. Futhermore, the pressure and the thermodynamic volume are considered associated with the cosmological constant and its conjugate variable, respectively~\cite{Kastor:2009wy}.
Based on the proposal, the pressure of AdS Black hole in extended phase space satisfies
\begin{equation}\label{eqn:Q31}
P=-\frac{1}{8\pi}\Lambda=\frac{(d-1)(d-2)}{16\pi l^{2}},
\end{equation}
where $d$ is interpreted as the dimensions of spacetime.

In \cite{Cvetic:2010jb}, Cvetic et al. raised the concept of reverse isoperimetric inequality and pointed out all the known black holes obey the equation, of which the expression is
\begin{equation}\label{eqn:Q33}
R=\left[\frac{(d-1)V}{\omega_{d-2}}\right]^{[1/(d-1)]}(\frac{\omega_{d-2}}{A})^{[1/(d-2)]}\geq1.
\end{equation}

When $R=1$, it implies the enthalpy of a black hole with confirmed thermodynamic volume reaches its upper bound.
Here, $A$ represents horizon area, $V$ signifies the thermodynamic volume mentioned above and $\omega_{d}$ measures the surface area
of the unit ball in $d+1$ dimensions expressed as $\omega_{d}=\frac{2\pi^{[(d+1)/2]}}{\Gamma[(d+1)/2]}
 $ in $d$-dimensional spacetime.
Nevertheless, the violations of the reverse isoperimetric inequality have been found in several black holes~\cite{{Hennigar:2014cfa},{Klemm:2014rda},{Hennigar:2015cja},{Brenna:2015pqa},{Song:2023kvq},{Noorbakhsh:2016faj},
{Noorbakhsh:2017tbp},{Feng:2017jub},{Noorbakhsh:2017nde},{Mann:2018jzt}}, whose entropies exceed the upper bound mentioned above, called super-entropy black hole.
However, futher investigation is impeded due to the lack of explanation of the phenomenon.

Generally speaking, the stability of the thermodynamic system is assessed by the
sign of $C_{P}$ and the negative one leads to instability in thermodynamics. Recently, with the volume term introduced to the first law of thermodynamics, a new suggestion assumes that the violation of isoperimetric inequality signifies thermodynamic instability, which denotes $C_{V}<0$.

The emergence of this hypothesis~\cite{Johnson:2019mdp} is based on that for a number of super-entropy black holes, $C_{V}$ is always negative.
Then, when researching a topical super-entropy black hole, Wan and Robert detected in certain range $C_{V}$ exceed zerobound while $C_{P}$ remain negative in this range~\cite{Cong:2019bud}.
Thus, they put forward a new proposal that the assessment of thermodynamic instability not only links to $C_{V}$, but also relates to the specific heat at constant pressure $C_{P}$.

Nevertheless, the conclusion is not deduced by enough black holes. For examining its universality, we reconsider the relation of $C_{V}$ and $C_{P}$ in another super-entropy black hole, the Kerr-Newman-AdS black hole in a coordinate system that rotates at infinity~\cite{Hennigar:2014cfa}. The concrete method to attained the black hole and the interesting properties will be mentioned below.

The rest of paper is organized as follows. In section~\ref{sec:A}, we review the thermodynamic
properties of Kerr-Newman AdS black holes. In section~\ref{sec:B}, we study and plot the curves of specific heat capacity. Finally, we conclude the results in section
~\ref{sec:C}.

\section{Kerr-Newman AdS black holes}
\label{sec:A}
In this section, we will introduce the thermodynamic properties of the Kerr-Newman AdS black holes in four-dimensional spacetime, of which the metric in the Boyer-Lindquist-like coordinates is written as
\begin{equation}\label{eqn:Q1}
ds^{2}=-\frac{\Delta_{r}}{\rho^{2}}(dt-\frac{asin^{2}\theta}{\Xi}d\phi)^{2}+\frac{\rho^{2}}{\Delta_{r}}dr^{2}+\frac{\rho^{2}}{\Delta_{\theta}}d\theta^{2}+\frac{sin^{2}\theta\Delta_{\theta}}{\rho^{2}}(adt-\frac{r^{2}+a^{2}}{\Xi}d\phi)^{2},
\end{equation}
where
\begin{equation}\label{eqn:Q2}
\rho^{2}=r^{2}+a^{2}cos^{2}\theta, \Xi=1-\frac{a^{2}}{l^{2}},\Delta_{r}=(r^{2}+a^{2})(1+\frac{r^{2}}{l^{2}})-2mr+q^{2},\Delta_{\theta}=1-\frac{a^{2}}{l^{2}}cos^{2}\theta.
\end{equation}

The roots of the $\Delta_{r}=0$ are the horizon radius of the black hole, where the largest one is $r_{+}$. The mass of the black hole can be expressed as
\begin{equation}\label{eqn:Q3}
M=\frac{(r_{+}^{2}+a^{2})(r_{+}^{2}+l^{2})+q^{2}l^{2}}{2r_{+}l^{2}\Xi^{2}}.
\end{equation}

 In order to avoid naked singularity, $r_{+}$ must satisfy the relation, which brings to new restrictions on the character of $M$, $J$ and $Q$, namely $M>\sqrt{J}$.
For Kerr-AdS black hole, this relation reduces to $2M^{2}>\sqrt{4J^{2}+Q^{4}}+Q^{2}$.

The horizon area $A$ in Eq.$\left(\ref{eqn:Q1}\right)$ reads
\begin{equation}\label{eqn:Q4}
A=\frac{-qr}{\rho}(dt-\frac{asin\theta}{\Xi}d\phi).
\end{equation}

As the entropy obeys relation $S=\frac{A}{4}$, the entropy is given by
\begin{equation}\label{eqn:Q5}
S=\frac{\pi(r_{+}^{2}+a^{2})}{\Xi}.
\end{equation}

Thus, utilizing Eq.$\left(\ref{eqn:Q1}\right)$ and Eq.$\left(\ref{eqn:Q2}\right)$, the Eq.$\left(\ref{eqn:Q3}\right)$ is recast as the new form related to the $S$,$Q$,$J$ and $P$
\begin{equation}\label{eqn:Q6}
M=\frac{1}{2}\sqrt{\frac{S}{\pi}[(1+\frac{\pi Q^{2}}{S}+\frac{8PS}{S^{2}})^{2}+\frac{4\pi^{2}J^{2}}{S^{2}}(1+\frac{8PS}{3})]}.
\end{equation}

Therefore, the first law of thermodynamics in extended spacetime changes into
\begin{equation}\label{eqn:Q32}
dM=TdS+VdP+\Omega dJ+\Phi dQ.
\end{equation}

Here, $J$ and $T$ are the angular momentum and Hawking temperature, while the $\Omega$
measures the angular velocity and $\Phi$ stands for the electric potential.
By means of the above formulas, these characters are given by
\begin{equation}\label{eqn:Q8}
\begin{aligned}
&T=(\frac{\partial M}{\partial S})_{P,J,Q}=\frac{1}{8\pi M}[(1+\frac{\pi Q^{2}}{S}+\frac{8PS}{3})(1-\frac{\pi Q^{2}}{S}+8PS)-\frac{4\pi^{2}J^{2}}{S^{2}}],\\
&V=(\frac{\partial M}{\partial P})_{S,J,Q}=\frac{2S^{2}}{3\pi M}(1+\frac{\pi Q^{2}}{S}+\frac{8PS}{3}+\frac{2\pi J^{2}}{S^{2}}),\\
&\Omega=(\frac{\partial M}{\partial J})_{S,P,Q}=\frac{\pi J}{MS}(1+\frac{8PS}{3}),\\
&\Phi=(\frac{\partial M}{\partial Q})_{S,P,J}=\frac{Q}{2M}(1+\frac{\pi Q^{2}}{S}+\frac{8PS}{3}).\\
\end{aligned}
\end{equation}

Finally, the pressure can be calculated from Smarr relation~\cite{Smarr:1972kt}
\begin{equation}\label{eqn:Q9}
M=2TS-2PV+2\Omega J+\Phi Q.
\end{equation}

Substituting Eq.$\left(\ref{eqn:Q32}\right)$ into Eq.$\left(\ref{eqn:Q8}\right)$, in this way we obtain a new expression of the variable $P$
\begin{equation}\label{eqn:Q10}
P=\frac{T}{v}-\frac{1}{2\pi v^{2}}+\frac{2Q^{2}}{\pi v^{4}}+\frac{48J^{2}}{\pi v^{6}},
\end{equation}
where $v=2[3V/(4\pi)]^{1/3}$ substitutes for the original volume for convenience.
\section{Thermodynamic instability}
\label{sec:B}
In~\cite{Hennigar:2014cfa}, Robie and Robert construct a new ultraspinning limit to the
spinning Kerr-AdS metric so that they obtain a new set of the solutions for black holes in four space-time dimensions. For a black hole in a coordinate system rotating at infinity, if we increase its speed to velocity of light and compactify the orientation of azimuth, the structure of the
space-time will change primarily a lot so that it won't come back to the initial frame. Then we obtain the new metric solutions which are tantamount to a set of solutions of black hole in gauged supergravity~\cite{{Klemm:2014rda},{Gnecchi:2013mja}}.
Through the method mentioned above, with noncompact horizon and restricted area, a super-entropy black hole emerges.
The specific steps are as follows.
Substituting the $\psi
 $ for $\phi/\Xi$ in metric and limiting $a\rightarrow l$, the metric of the Kerr-Newman-AdS is reconsidered as
\begin{equation}\label{eqn:Q11}
\begin{aligned}
&ds^{2}=-\frac{\Delta}{\Sigma}[dt-lsin^{2}\theta d\psi]^{2}+\frac{\Sigma}{\Delta}dr^{2}+\frac{\Sigma}{sin^{2}\theta}d\theta^{2}+\frac{sin^{4}\theta}{\Sigma}[ldt-(r^{2}+l^{2})d\psi]^{2},\\
&A=-\frac{qr}{\Sigma}(dt-lsin^{2}\theta d\psi),\\
\end{aligned}
\end{equation}
where
\begin{equation}\label{eqn:Q12}
\Sigma=r^{2}+l^{2}cos^{2}\theta
\Delta=(l+\frac{r^{2}}{l})^{2}-2mr+q^{2}.
\end{equation}

The elementary thermodynamic quantities of the
super-entropy black hole are given by
\begin{equation}\label{eqn:Q13}
\begin{aligned}
&M=\frac{\mu m}{2\pi},
J=Ml,
\Omega=\frac{l}{r_{+}^{2}+l^{2}},\\
&T=\frac{1}{4\pi r_{+}}(3\frac{r_{+}^{2}}{l^{2}}-1-\frac{q^{2}}{r_{+}^{2}+l^{2}}),
S=\frac{u}{2}(r_{+}^{2}+l^{2})=\frac{A}{4},\\
&\Phi=\frac{qr_{+}}{r_{+}^{2}+l^{2}},
Q=\frac{uq}{2\pi}.\\
\end{aligned}
\end{equation}

These parameters are computed by the method of conformal completion in~\cite{{Ashtekar:1984zz},{Ashtekar:2000zz},{Das:2000cu}} with the killing vectors $\partial_{\tau}$
 and $\partial_{\psi}$ as well as the $"$chirality~condition$"$ $J=ML$.

\begin{figure}[htb]
\label{all1}
\begin{center}
\vspace{-0.5cm}
\subfigbottomskip=1pt
\subfigcapskip=3pt
\subfigure[$0<r_{+}<2$]{
\includegraphics[width=0.6\textwidth]{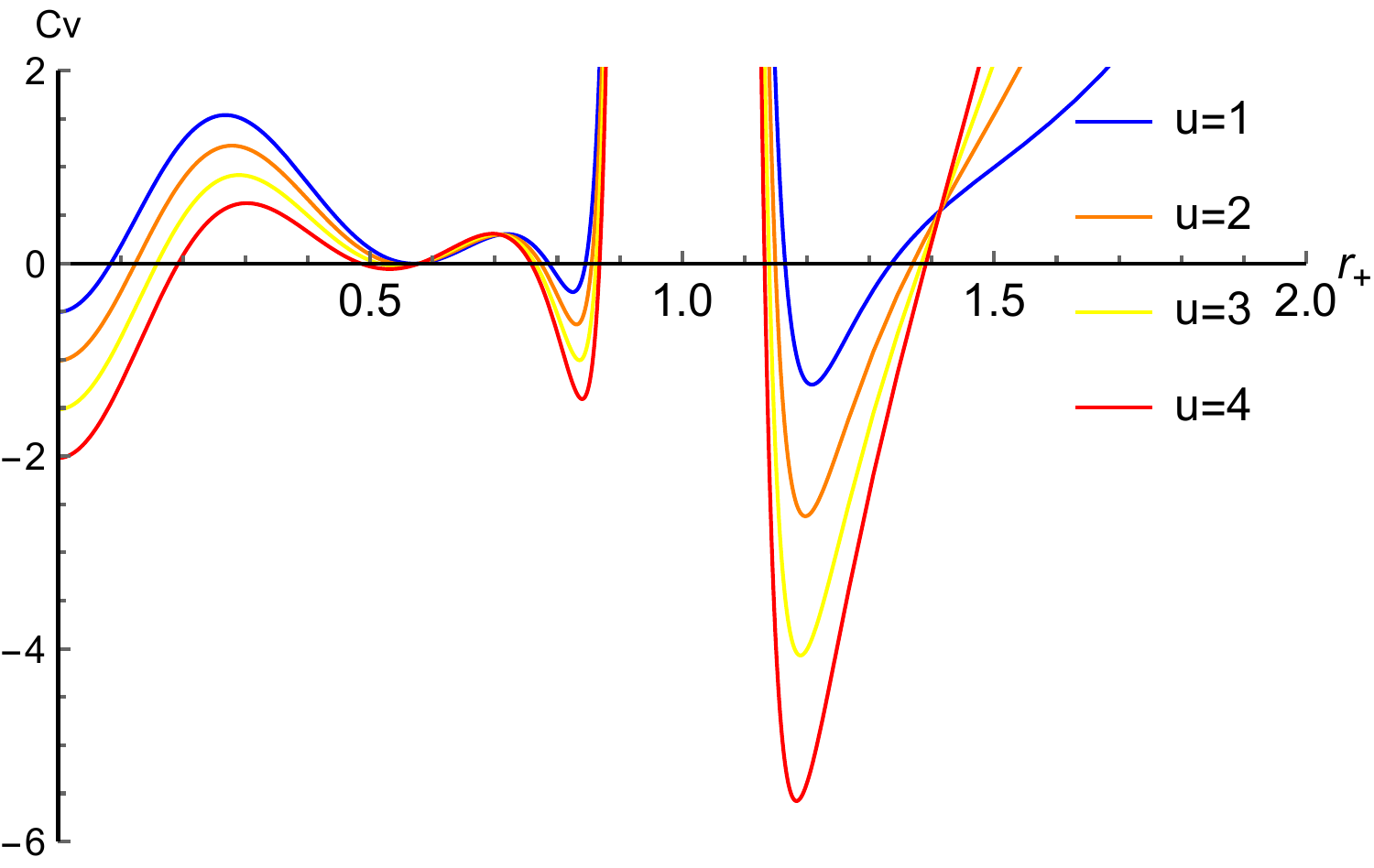}}
\label{fig:a01}

\subfigure[$0.5<r_{+}<0.8$]{
\includegraphics[width=0.6\textwidth]{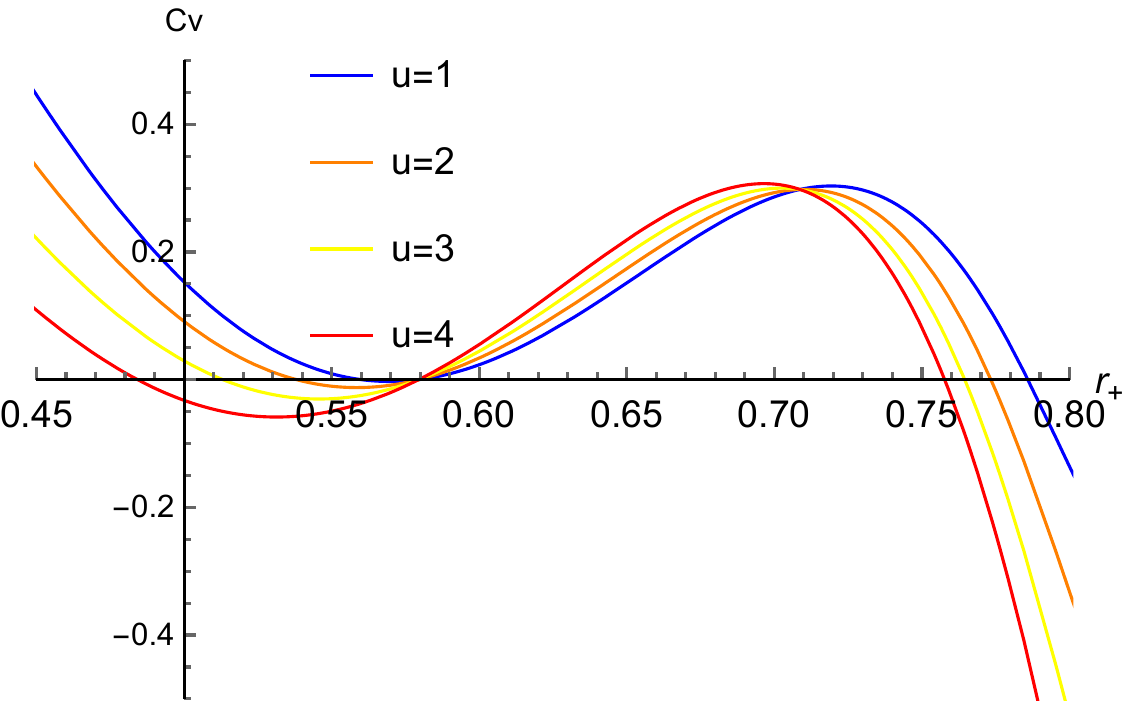}}
\label{fig:a02}

\end{center}
\caption{$C_{V}$ for different $u$ of Kerr-Newman-AdS black hole, $q=0.1,l=1$}
\label{fig:1}
\end{figure}
So as to investigate the thermodynamic stability of the Kerr-Newman-AdS black hole, we consider \begin{equation}\label{eqn:Q20}
C_{P}=T\frac{\partial S}{\partial T}|_{P,J}.
\end{equation}

 By utilizing Eq.$\left(\ref{eqn:Q5}\right)$ and Eq.$\left(\ref{eqn:Q32}\right)$,
$C_{V}$ is calculated by the follow equation
\begin{equation}\label{eqn:Q21}
C_{V}=C_{P}-TV\alpha_{T}^{2}\kappa_{T},
\end{equation}
where $\alpha_{T}=
 V^{-1}\frac{\partial V}{\partial T}|_{P,J}$
  and $\kappa_{T}=-V\frac{\partial P}{\partial V}|_{T,J}$.
 Finally, replacing $S$ with expression of $r$ in Eq.$\left(\ref{eqn:Q20}\right)$ and Eq.$\left(\ref{eqn:Q21}\right)$ by formula  Eq.$\left(\ref{eqn:Q5}\right)$, we receive the resultant expressions of $C_{P}$ and $C_{V}$. Considering the expressions are too long, we exclude the concrete expressions here.

\begin{figure}[htb]
\label{all1}
\begin{center}
\vspace{-0.5cm}
\subfigbottomskip=1pt
\subfigcapskip=3pt
\subfigure[$0<r_{+}<2$]{
\includegraphics[width=0.6\textwidth]{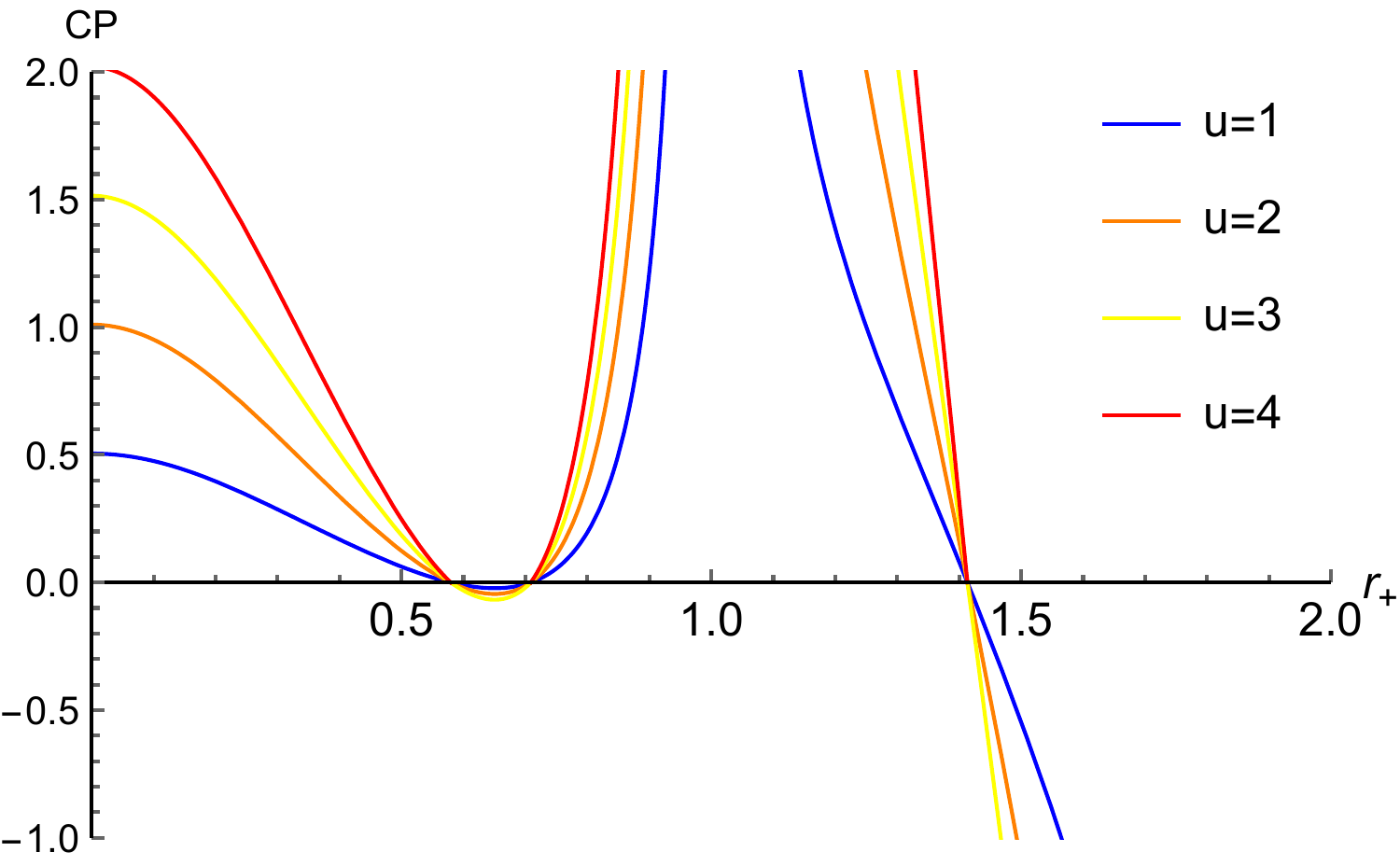}}
\label{fig:a01}

\subfigure[$0.5<r_{+}<0.8$]{
\includegraphics[width=0.6\textwidth]{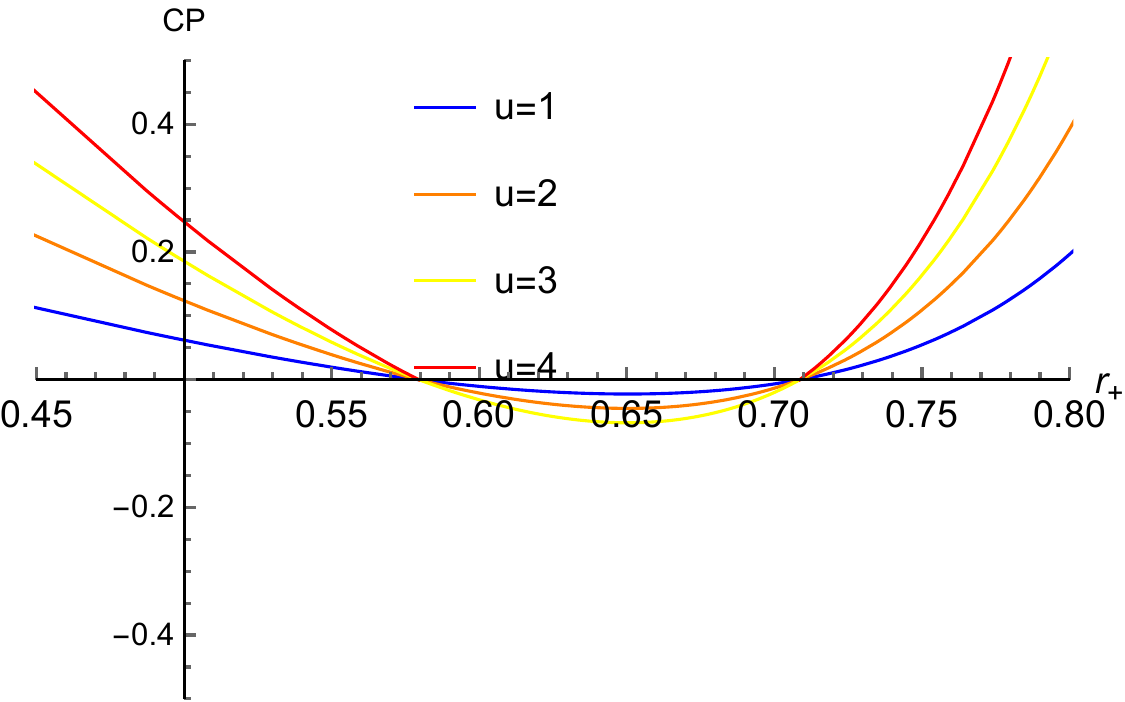}}
\label{fig:a02}

\end{center}
\caption{$C_{P}$ for different $u$ of Kerr-Newman-AdS black hole, $q=0.1,l=1$}
\label{fig:2}
\end{figure}

The figures of various $u$ gathered in Fig.~\ref{fig:1}. To obtain the more detailed information, we choose the smaller range to plot other figures placed the original figure below. The figures imply an interesting phenomenon that when $r_{+}$ is less than a certain value, $C_{V}$ will exceed zero bound,
inconsistent with~\cite{Johnson:2019mdp}. In~\cite{Cong:2019bud}, Wan and Robert noticed where $C_{V}>0$, meanwhile $C_{P}<0$ for the super-entropy black hole investigated. As $C_{P}<0$ indicates the instability of thermodynamic system, which hints that the super-entropy black hole is unstable.
Considering the assumption in \cite{Cong:2019bud}, we plot the figure related to  $C_{P}$ in Fig.~\ref{fig:2}. Obviously, the range $C_{V}>0$ in Fig.~\ref{fig:1} and $C_{P}<0$ in Fig.~\ref{fig:2} are not in coincidence, which proposes a counterexample
to the conjecture in~\cite{Cong:2019bud}. Besides, as parameter $u$ increases, the null points of $C_{P}$ and $C_{V}$ shift right, which indicates $u$ impacts on the thermodynamic instability of the system for the Kerr-Newman-AdS black hole in a coordinate system that rotates at infinity.

\section{Conclusion}
\label{sec:C}
In this paper, we have proved that the Kerr-Newman-AdS black hole in a coordinate system that rotates at infinity violates the conjecture in \cite{Johnson:2019mdp} because the black hole disobeys the reverse isoperimetric inequality while $C_{V}$ is not always negative. Meanwhile, in~\cite{Cong:2019bud}, Wan and Robert noticed that $C_{P}<0$ when $C_{V}>0$. Therefore, the further assumption come forward that the black hole disobeys the reverse isoperimetric inequality will be erratic in thermodynamics. However, we find a counterexample
to Wan et al. that the region $C_{V}<0$ and $C_{P}>0$ are not in coincidence completely. Observing the null point of $C_{V}$ and $C_{P}$, we inferred the thermodynamic instability of the system influenced by the variable $u$ for the black hole investigated. For other super-entropy black holes, perhaps there will be the similar parameter affecting the thermodynamic instability of the system.
\section{ACKNOWLEDGEMENTS}
We are grateful to Yue Song, Deyou Chen, Rui Yin, Jing Liang, Peng Wang, Haitang Yang, Jun Tao and Xiaobo Guo for useful discussions. This work is supported in part by NSFC (Grant No. 11747171), Xinglin Scholars Project of Chengdu University of Traditional Chinese Medicine (Grant no.QNXZ2018050).

\end{document}